\documentclass[a4paper,11pt]{article}
\usepackage{pos}

\usepackage{graphicx}
\usepackage{media9}

\usepackage{tikz,varwidth}
\usetikzlibrary{shapes.geometric}
\usetikzlibrary{arrows.meta,shapes,automata,petri,positioning,calc,decorations.markings}

\title{Impact of Dynamical Fermions on Centre Vortex Structure}

\author*[a]{Derek Leinweber}
\author[a]{James Biddle}
\author[a]{Waseem Kamleh}

\affiliation[a]{Centre for the Subatomic Structure of Matter, Department of Physics,\\ The University of Adelaide, SA 5005, Australia}

\emailAdd{derek.leinweber@adelaide.edu.au}
\emailAdd{james.biddle@adelaide.edu.au}
\emailAdd{waseem.kamleh@adelaide.edu.au}

\abstract{This presentation examines the centre vortex structure of Monte-Carlo generated
  gauge-field configurations using modern visualisation techniques.  This time, the manner in which
  light dynamical fermion degrees of freedom impact the centre-vortex structure is explored.
  Focusing on the thin vortices identified by plaquettes having a non-trivial centre phase, the
  vortex structure is illustrated through 3D renderings of oriented plaquettes.
  The impact of light dynamical fermions is not subtle, changing both the density of vortices and
  the complexity of the vortex structures observed.  The role of vortex branching points in full
  QCD is highlighted in the survey of results presented.  }

\FullConference{%
 The 38th International Symposium on Lattice Field Theory, LATTICE2021
  26th-30th July, 2021
  Zoom/Gather@Massachusetts Institute of Technology
}

\begin{document}
\maketitle

\section{Introduction}

The salient emergent features of the QCD ground-state vacuum fields
are: the dynamical generation of mass through chiral symmetry
breaking, and the confinement of quarks.  Understanding the most
fundamental aspects of QCD-vacuum field structure giving rise to these
phenomena is a contemporary problem of interest.

The most fundamental mechanism proposed to give rise to these
phenomena are centre vortices
\cite{tHooft:1977nqb,tHooft:1979rtg,DelDebbio:1996lih,Faber:1997rp,%
  DelDebbio:1998luz,Bertle:1999tw,Faber:1999gu,Engelhardt:1999xw,%
  Bertle:2000qv,Greensite:2003bk,Engelhardt:2003wm,Greensite:2016pfc}
within the ground-state vacuum fields.  Here, the eight degrees of
freedom contained within an $SU(3)$ gauge-field link are replaced by
one of the three cube roots of unity.  It is this extreme level of
simplification that makes the centre-vortex proposal so compelling as
the most fundamental essence of QCD vacuum structure.

Within the pure-gauge sector of $SU(3)$-colour, there is encouraging
evidence that centre vortices underpin both confinement and dynamical
chiral-symmetry breaking.  
\begin{itemize}
\item Removal of $SU(3)$ centre vortices removes confinement, while
  consideration of the vortices alone provides confinement
 ~\cite{Langfeld:2003ev,Cais:2008za, Trewartha:2015ida}.

\item The planar vortex density of centre-vortex degrees of freedom
  scales with the lattice spacing providing an well defined continuum
  limit~\cite{Langfeld:2003ev}.

\item Removal of vortices suppresses the infrared enhancement of the
  gluon propagator.  Again the vortices alone contain the long
  distance structure of the gluon fields responsible for the
  well-known infrared enhancement~\cite{Biddle:2018dtc}.

\item A connection between centre vortices and instantons was
  established through gauge-field smoothing~\cite{Trewartha:2015ida}.
  An understanding of the phenomena linking these degrees of
  freedom was illustrated in Ref.~\cite{Biddle:2019gke}.

\item Centre vortices have been shown to give rise to mass splitting
  in the low-lying hadron spectrum
 ~\cite{Trewartha:2017ive,Trewartha:2015nna,OMalley:2011aa}.

\item Through studies of the nonperturbative quark propagator of the
  overlap-Dirac fermion operator, evidence that centre vortices
  underpin dynamical chiral symmetry breaking in $SU(3)$ gauge theory
  was reported in Ref.~\cite{Trewartha:2015nna}.

\item The removal of centre vortex degrees of freedom from the gluon
  fields restores chiral symmetry~\cite{Trewartha:2017ive}.
\end{itemize}

But how do these results hold up in QCD where the virtual transition
of gluons into dynamical quark-antiquark pairs can significantly alter
the vacuum structure~\cite{Moran:2008qd}.  In this presentation we
investigate how these dynamical fermions change the centre-vortex
structure of the ground-state vacuum fields.

\section{Centre Vortex Identification}

Centre vortices are identified through a gauge fixing procedure
designed to bring the lattice link variables as close as possible to
the identity, up to a phase.  Here, we consider the original
Monte-Carlo generated configurations and gauge transform them directly
to Maximal Centre Gauge
\cite{DelDebbio:1996mh,Langfeld:1997jx,Langfeld:2003ev}, avoiding any
preconditioning~\cite{Cais:2008za}.  In doing so, the lattice link
variables $U_\mu(x)$ are brought close to the centre elements of
$SU(3)$,
\begin{equation} 
Z = \exp \left ( 2 \pi i\, \frac{m}{3} \right ) \, \mathbf{I}, \textrm{ with } m = -1, 0, 1.
\label{CentreSU3}
\end{equation}
This is implemented through gauge transformations $\Omega$ such that,
\begin{equation}
\sum_{x,\mu} \,  \left | \mathrm{tr}\, U_\mu^\Omega(x) \, \right |^2 \stackrel{\Omega}{\to}
\mathrm{max} \, .
\label{GaugeTrans}
\end{equation}
One then projects the link variables to the centre 
\begin{equation} 
U_\mu(x) \to Z_\mu(x) \textrm{ where }
Z_\mu(x) = \exp \left ( 2 \pi i\, \frac{m_\mu(x)}{3} \right
)\mathbf{I} \, ,
\label{UtoZ}
\end{equation}
where $m_\mu(x) = -1, 0, 1.$

The centre-line of an extended vortex in three dimensions is
identified by the presence of nontrivial centre charge, $z$, found in
the product of centre-projected links around a plaquette,
\begin{equation}
z = \prod_\Box Z_\mu(x) = \exp \left ( 2 \pi i\, \frac{m}{3} \right ) \, .
\label{CentreCharge}
\end{equation}
A right-handed ordering of the dimensions is selected in calculating
the centre charge.  If $z=1$, no vortex pierces the plaquette. If $z
\neq 1$ a vortex with charge $z$ pierces the plaquette.  We refer to
the centre charge of a vortex via the value of $m = \pm 1$.

\section{Centre Vortex Visualisation}

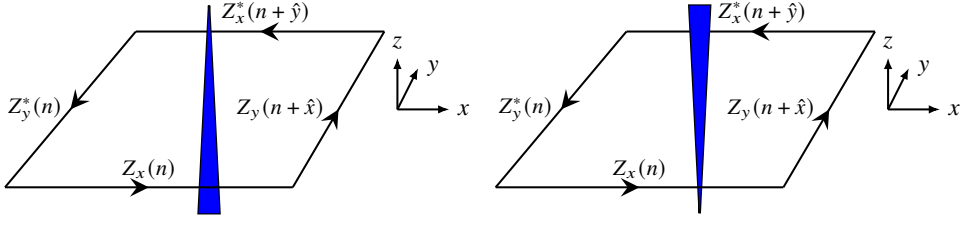
\begin{figure}[t]
\begin{center}
\resizebox{!}{3cm}{%
\begin{tikzpicture}[scale=0.9]
\begin{scope}[very thick,decoration={
    markings,
    mark=at position 0.5 with {\arrow[scale=2]{stealth}}}
    ] 
  \draw[line width=1.0,postaction={decorate}](1.5,-1.5)-- node[left]{$Z_y(n+\hat x)\ $} (3.25,1.5)node(g){};
  \draw[line width=1.0,postaction={decorate}](3.25,1.5)-- node[above]{${}\ \ Z_x^*(n+\hat y)$} (-1.5,1.5);
  \draw[line width=1.0,postaction={decorate}](-1.5,1.5)-- node[left]{$Z_y^*(n)$}(-4,-1.5);

  \draw (-0.3,-2) node(a){}
  -- (0.1,-2) node(b){}   
  -- (-0.1,2) node(c){}   
  -- cycle;               
  \fill[blue] (a.center) -- (b.center) -- (c.center);
  
  \draw[line width=1.0,postaction={decorate}](-4,-1.5)-- node[above]{$Z_x(n)$}(1.5,-1.5)node(f){};
  
  \draw[line width=1.0,-{Latex[length=2mm]}](3.5,0)--(4.5,0.0)node[right]{\large $x$};
  \draw[line width=1.0,-{Latex[length=2mm]}](3.5,0)--(3.9,0.8)node[right]{\large $y$};
  \draw[line width=1.0,-{Latex[length=2mm]}](3.5,0)--(3.5,1.0)node[above]{\large $z$};
  \end{scope}
\end{tikzpicture}
}
\resizebox{!}{3cm}{%
\begin{tikzpicture}[scale=0.9]
\begin{scope}[very thick,decoration={
    markings,
    mark=at position 0.5 with {\arrow[scale=2]{stealth}}}
    ] 
  \draw[line width=1.0,postaction={decorate}](1.5,-1.5)-- node[left]{$Z_y(n+\hat x)\ $} (3.25,1.5)node(g){};
  \draw[line width=1.0,postaction={decorate}](3.25,1.5)-- node[above]{\quad $Z_x^*(n+\hat y)$} (-1.5,1.5);
  \draw[line width=1.0,postaction={decorate}](-1.5,1.5)-- node[left]{$Z_y^*(n)$}(-4,-1.5);

  \draw (-0.3,2) node(a){}
  -- (0.1,2) node(b){}
  -- (-0.1,-2)node(c){}
  -- cycle;
  \fill[blue] (a.center) -- (b.center) -- (c.center);
  
  \draw[line width=1.0,postaction={decorate}](-4,-1.5)-- node[above]{$Z_x(n)$}(1.5,-1.5)node(f){};
  
  \draw[line width=1.0,-{Latex[length=2mm]}](3.5,0)--(4.5,0.0)node[right]{\large $x$};
  \draw[line width=1.0,-{Latex[length=2mm]}](3.5,0)--(3.9,0.8)node[right]{\large $y$};
  \draw[line width=1.0,-{Latex[length=2mm]}](3.5,0)--(3.5,1.0)node[above]{\large $z$};
  \end{scope}
\end{tikzpicture}
}
\end{center}
\vspace{-18pt}
\caption{{\bf Rendering the centre charge} of Eq.~(\protect\ref{CentreCharge}) associated with a
  plaquette in the $x$-$y$ plane at lattice site $n$.  (left) An $m = +1$ vortex with centre charge
  $z = \exp(2\pi i / 3)$ is rendered as a jet pointing in the $+\hat z$ direction.  (right) An
  $m = -1$ vortex with centre charge $z = \exp(-2\pi i / 3)$ is rendered as a jet in the $-\hat
  z$ direction.  }
\label{fig:jets}
\end{figure}

The centre lines of extended vortices are illustrated on the dual
lattice by a jet piercing the plaquette producing a nontrivial centre
charge.  The orientation of the jet follows a right-handed coordinate
system.  For example, with reference to Eq.~(\ref{CentreCharge}), an
$m = +1$ vortex in the $x$-$y$ plane is plotted in the $+\hat z$
direction as a blue jet.  Similarly, an $m = -1$ vortex in the $x$-$y$
plane is plotted in the $-\hat z$ direction.  Figure \ref{fig:jets}
provides an illustration of this assignment.

As the centre charge transforms to its complex conjugate under permutation of the two
dimensions describing the plaquette, the centre charge can be thought of as a directed flow of
charge $z = \exp(2\pi i / 3)$.  

\section{Centre Vortex Structure}

The projected centre vortices (P-vortices), identified on the lattice
as described above, form surfaces in four dimensional space-time,
analogous to the centre line of a vortex in fluid dynamics that maps
out a surface as it moves through time.  As the surface cuts through
the three-dimensional volume of our visualisation, a P-vortex line
mapping the flow of centre charge is rendered.

Our point of focus in this presentation is to discover the impact of
dynamical-fermion degrees of freedom on the vortex structure of a
gauge field.  Here we the PACS-CS $(2+1)$-flavour full-QCD
ensembles~\cite{Aoki:2008sm}, made available through the
ILDG~\cite{Beckett:2009cb}. These ensembles use a $32^3 \times 64$
lattice, and employ a renormalisation-group improved Iwasaki gauge
action with $\beta = 1.90$ and non-perturbatively ${\cal
  O}(a)$-improved Wilson quarks, with $C_{\rm SW} = 1.715$.  We
consider their lightest $u$- and $d$-quark-mass ensemble identified by
a pion mass of 156 MeV~\cite{Aoki:2008sm}, and set the scale using the
Sommer parameter with $r_0 = 0.4921$ fm providing a lattice spacing of
$a=0.0933$ fm~\cite{Aoki:2008sm}.  

For comparison, we have created a matched $32^3 \times 64$ pure-gauge
ensemble using the same improved Iwasaki gauge action with $\beta =
2.58$ providing a Sommer-scale spacing of $a = 0.100$ fm, facilitating
comparisons with all the PACS-CS ensembles.

\begin{figure}[p]
\null\hspace{-0.3cm}
\includemedia[
        noplaybutton,
	3Dtoolbar,
	3Dmenu,
	3Dviews=U3D/Primary-Secondary-PG-28.vws,
	3Dcoo  = 16 16 32, 
        3Dc2c=0.245114266872406 0.8673180341720581 0.43321868777275085,
	3Droo  = 110.0,    
	3Droll =-98.1,     
	3Dlights=Default,  
	width=1.04\textwidth,  
]{\includegraphics{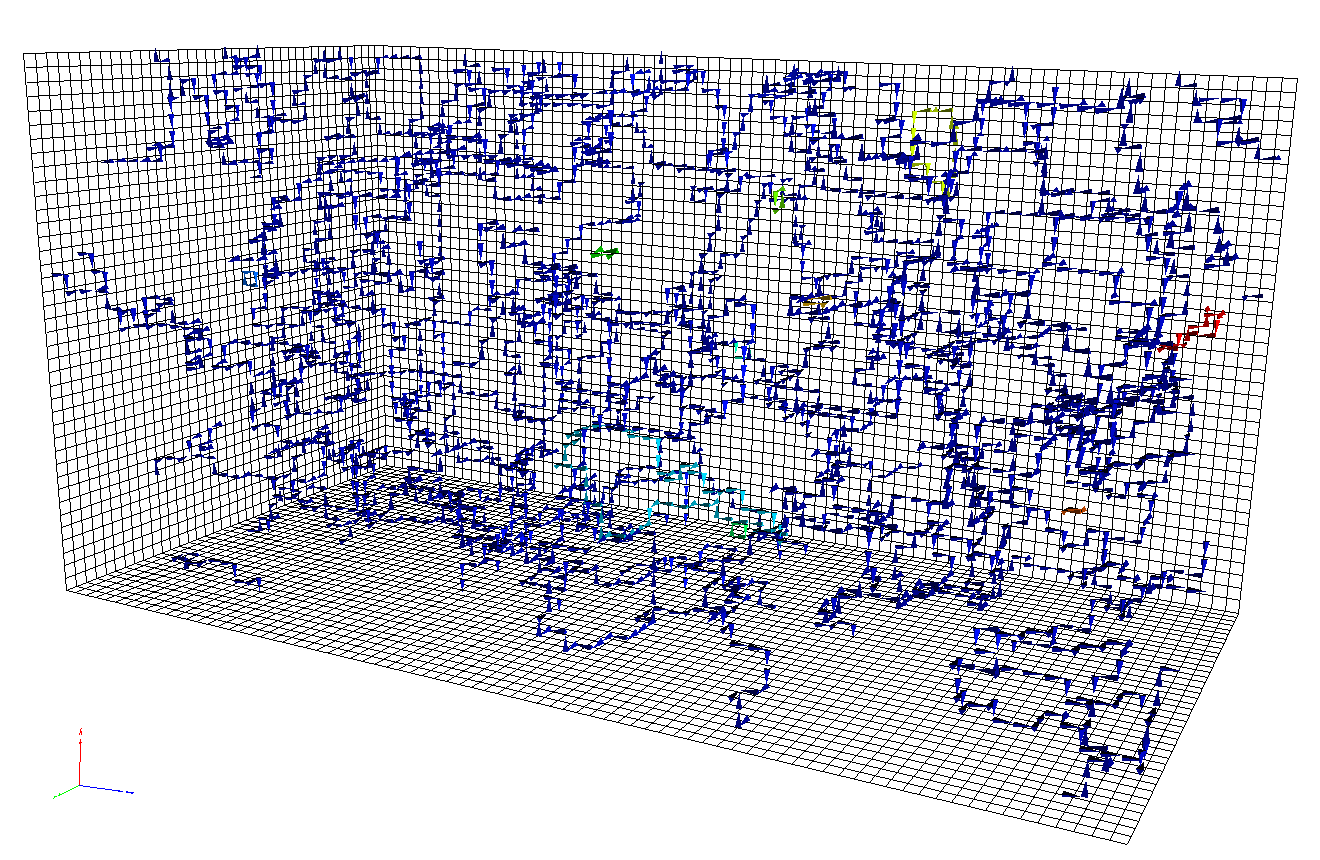}}{U3D/Primary-Secondary-PG-28.u3d}
\vspace{-24pt}
\caption{
  {\bf The centre-vortex structure of a ground-state vacuum field
    configuration in pure SU(3) gauge theory.}
  ({\sl Click to activate.})
  The flow of $+1$ centre charge through a gauge field is illustrated by the jets.  
  Blue jets illustrate the single percolating vortex structure, while
  other colours illustrate smaller structures.
\label{Primary-Secondary-PG-28.u3d}
}
\begin{center}
\includemedia[
        noplaybutton,
	3Dtoolbar,
	3Dmenu,
	3Dviews=U3D/Primary-Secondary-DF-18.vws,
	3Dcoo  = 16 16 32, 
        3Dc2c=0.245114266872406 0.8673180341720581 0.43321868777275085,
	3Droo  = 110.0,    
	3Droll =-98.1,     
	3Dlights=Default,  
	width=\textwidth,  
]{\includegraphics{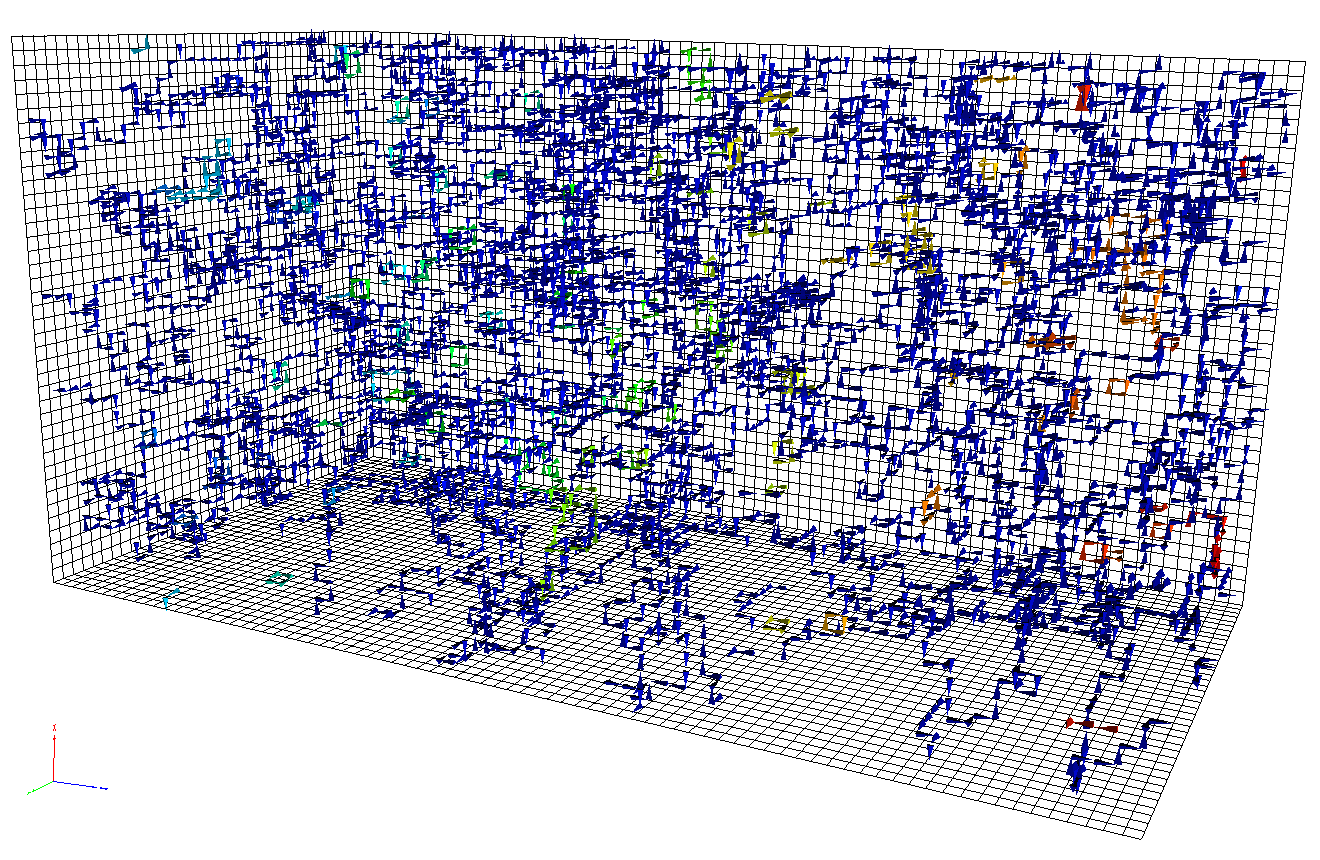}}{U3D/Primary-Secondary-DF-18.u3d}
\end{center}
\vspace{-18pt}
\caption{
  {\bf The centre-vortex structure of a ground-state vacuum field
    configuration in dynamical 2+1 flavour QCD.}
  ({\sl Click to activate.})
  The flow of $+1$ centre charge through a gauge field is illustrated by the jets.  
  Symbols are as described in Fig.~\ref{Primary-Secondary-PG-28.u3d}.
\label{Primary-Secondary-DF-18.u3d}
}
\end{figure}

Figures \ref{Primary-Secondary-PG-28.u3d} and
\ref{Primary-Secondary-DF-18.u3d} illustrate the centre-vortex
structure of pure-gauge and dynamical-fermion ground-state vacuum
fields respectively.  Inspection of the vortices reveals the flow of
centre charge, intersection points and a prevalence of branching
points resembling monopole or anti-monopole contributions, where three
jets emerge from or converge to a point.  With the introduction of
dynamical fermions the structure becomes more complicated, both in the
abundance of nontrivial centre charge and in the proliferation of
branching points.

Figures \ref{Primary-Secondary-PG-28.u3d} and
\ref{Primary-Secondary-DF-18.u3d} are interactive illustrations which
can be activated in Adobe Reader\footnote{Open this pdf document in
  Adobe Reader 9 or later.  Linux users can install
  \href{ftp://ftp.adobe.com/pub/adobe/reader/unix/9.x/9.4.1/enu/}{Adobe
    acroread version 9.4.1}, the last edition to have full 3D support.
  From the ``Edit'' menu, select ``Preferences...'' and ensure ``3D \&
  Multimedia'' is enabled and ``Enable double-sided rendering'' is
  selected.}  by clicking on the image.  Once activated, click and
drag to rotate, Ctrl-click to translate, Shift-click or mouse wheel to
zoom, and right click to access the ``Views'' menu.  Several views
have been created to facilitate and inspection of the centre-vortex
structure.
The presence of a percolating vortex cluster is a characteristic
feature of the confining phase~\cite{Engelhardt:1999fd}.  These
illustrations are representative of the ensemble in that the vortex
vacuum is typically dominated by a single large percolating cluster.
Moreover, dynamical fermions significantly increase the number of
vortices observed.

Considering an ensemble of 200 configurations with $32$
three-dimensional slices each, the average number of vortices
composing the primary cluster in these $32^2 \times 64$ spatial slices
is
\begin{itemize}
\item $3,277\, \pm 156$ vortices in the Pure Gauge theory, versus
\item $5,924\, \pm 239$ vortices in Full QCD.
\end{itemize}
Noting that there are $32^2 \times 64 \times 3 = 196,608$ spatial
plaquettes on these lattices, one sees that the presence of a vortex
is still a relatively rare occurrences.

\begin{figure}[p]
\begin{center}
\includemedia[
        noplaybutton,
	3Dtoolbar,
	3Dmenu,
	3Dviews=U3D/Secondary-PG-28.vws,
	3Dcoo  = 16 16 32, 
        3Dc2c=0.245114266872406 0.8673180341720581 0.43321868777275085,
	3Droo  = 110.0,    
	3Droll =-98.1,     
	3Dlights=Default,  
	width=1.04\textwidth,  
]{\includegraphics{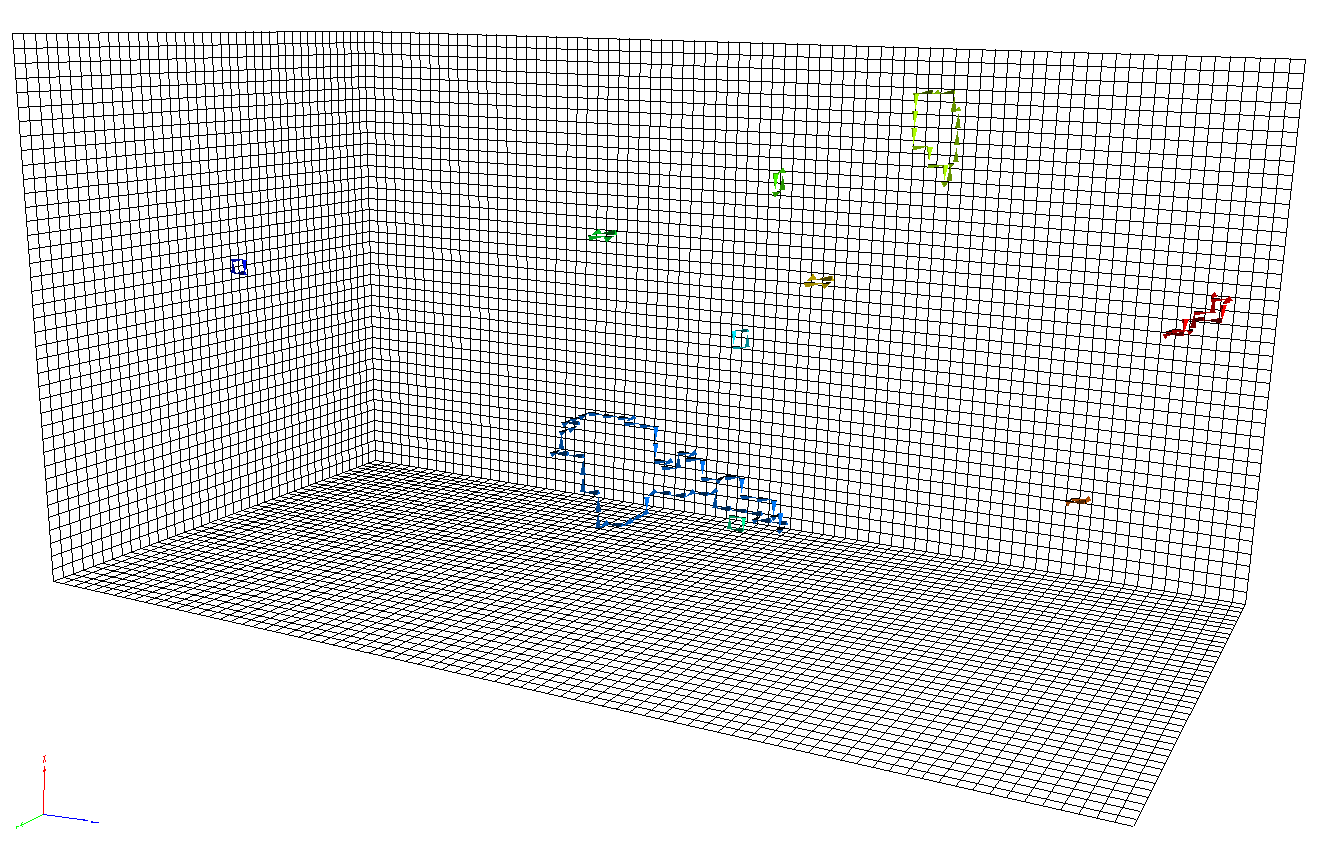}}{U3D/Secondary-PG-28.u3d}
\end{center}
\vspace{-24pt}
\caption{
  {\bf The centre-vortex structure of secondary loops in a ground-state vacuum field
    configuration of pure SU(3) gauge theory.}
  ({\sl Click to activate.})
  The flow of $+1$ centre charge in the secondary loops -- left behind as the
  single percolating structure is removed -- is illustrated.
  %
\label{Secondary-PG-28.u3d}
}
\null\hspace{-0.25cm}
\includemedia[
        noplaybutton,
	3Dtoolbar,
	3Dmenu,
	3Dviews=U3D/Secondary-DF-18.vws,
	3Dcoo  = 16 16 32, 
        3Dc2c=0.245114266872406 0.8673180341720581 0.43321868777275085,
	3Droo  = 110.0,    
	3Droll =-98.1,     
	3Dlights=Default,  
	width=1.05\textwidth,  
]{\includegraphics{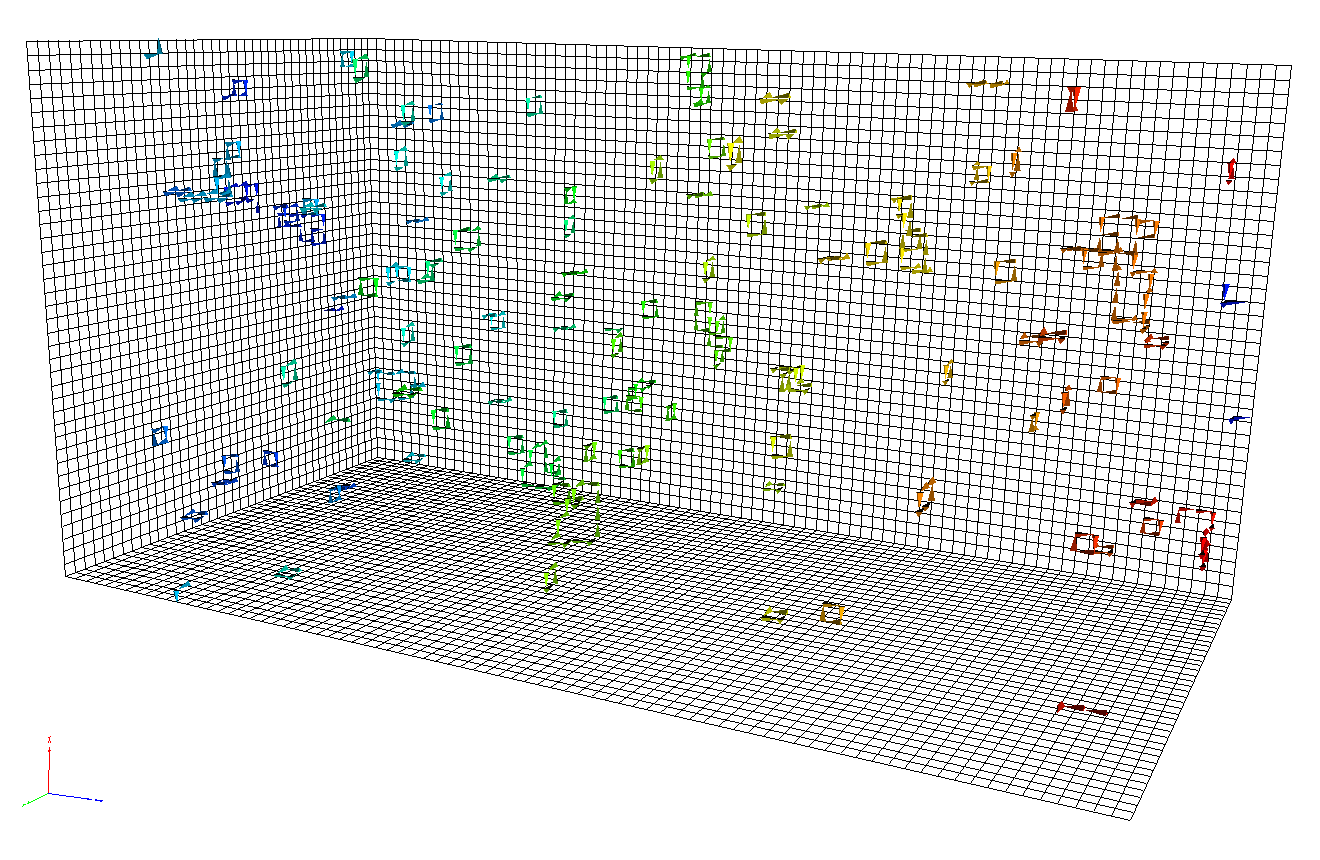}}{U3D/Secondary-DF-18.u3d}
\vspace{-24pt}
\caption{
  {\bf The centre-vortex structure of secondary loops in a ground-state vacuum field
    configuration of dynamical 2+1 flavour QCD.}
  ({\sl Click to activate.})
  The flow of $+1$ centre charge in the secondary loops is
  illustrated.
  %
\label{Secondary-DF-18.u3d}
}
\end{figure}

Figures \ref{Secondary-PG-28.u3d} and \ref{Secondary-DF-18.u3d}
illustrate the secondary loop structure left behind as one removes the
single large percolating structure.  The introduction of dynamical
fermions increases both the number of loops observed, and the
complexity of their structure by a proliferation of branching points
(or monopoles~\cite{Spengler:2018dxt}).  Figure
\ref{Secondary-DF-18.u3d} contains many views to facilitate the
observation of this complexity.

\section{Outlook}

In summary,
\begin{itemize}
\addtolength{\itemsep}{2pt}
\item Centre-vortex structure is complex.
\item We observe a proliferation of branching points in $SU(3)$ gauge
  theory with further enhancement as light dynamical fermion degrees
  of freedom are introduced in simulating QCD.
\item Each ground-state configuration is dominated by a long-distance
  percolating centre-vortex structure.
\item We have observed an approximate doubling in the size of the percolating
  vortex structure in in going from pure-gauge theory to full QCD.
\item We have also observed an enhancement in the number of small
  vortex paths upon the introduction of dynamical fermions.
\item Increased complexity in the vortex paths is also observed as the
  number of  monopole-antimonopole pairs is significantly increased
  with the introduction of dynamical fermions.
\end{itemize}
In short, dynamical-fermion degrees of freedom radically alter the
centre-vortex structure of the ground-state vacuum fields.  

Having gained insight into the impact of dynamical fermions on
ground-state vacuum field structure, future work will aim to quantify
these effects.  Here the distribution of path lengths is of particular
interest as it may be possible to quantify the change in the branching
probability of vortex paths in $SU(3)$ gauge theory as dynamical
fermions are introduced.

\section*{Acknowledgements}

We thank the PACS-CS Collaboration for making their $2+1$ flavour configurations available and the
ongoing support of the International Lattice Data Grid (ILDG).
This research is supported with supercomputing resources provided by the Phoenix HPC service at the
University of Adelaide and the National Computational Infrastructure (NCI) supported by the
Australian Government. This research is supported by the Australian Research Council
through Grants No.\ LE190100021, DP190102215, and DP210103706.
WK is supported by the Pawsey Supercomputing Centre through the Pawsey Centre for Extreme Scale Readiness (PaCER) program.

\bibliography{ImpactDynFerm}


\end{document}